**Four qubits Hamiltonian of the *Rs. molischianum* light-harvesting complex II ring**

Samir Lipovaca
slipovaca@nyse.com

*Abstract*: We will construct a simple four qubits Hamiltonian of the *Rs. molischianum* purple bacteria light harvesting complex II (LH-II) ring which yields energy levels that carry the ring's oscillator strength. In an excitonic representation, these levels are associated with the second and third lowest electronic excitations of the ring. We will assume that qubits form a closed loop lattice and the interaction between qubits is only due to the exchange effect. As we will show, eigenstates are constructed in a such way that as we subsequently divide qubits of the *Rs. molischianum* LH-II ring into the subsystem A consisting of only one qubit and the subsystem B consisting of the remaining qubits, respective entropies of entanglement increase until the value of 1 for the maximally entangled state (bipartite system) is reached. Since the Hamiltonian in essence introduces a two-level approximation for the LH-II ring, we will go one step further and assume that interactions between qubits closed loop lattices and an electromagnetic field are described by the Jaynes-Cummings Hamiltonian. This assumption is interesting by itself, since it leads to behavior where the qubits lattice and field oscillate back and fourth exchanging a quantum of energy, at the Rabbi frequency. This opens a challenging opportunity to experimentally study the *Rs. molischianum* LH-II ring in the regime of cavity QED.

## (1) Introduction

In the previous paper [1] we were shown by suitable labeling of the lowest energy ($Q_y$) excited states of individual bacteriochlorophylls (BChls) in the light-harvesting complexes II (LH-II) of purple bacteria *Rs. molischianum* and *Rps. acidophila* that the resulting exciton states are the quantum Fourier transform of the BChls excited states. For the *Rs. molischianum* LH-II ring exciton states were modeled as the four qubits quantum Fourier transform and we derived an explicit quantum circuit. In this paper we will present a simple Hamiltonian for the four qubits of the *Rs. molischianum* LH-II ring.
It was shown in [2] that exciton state energies are given as

$$E_n = E_0 + 2V_0 \cos\frac{n\pi}{N}, \quad n = -N+1, -N+2, \ldots, N \tag{1}$$

where $N = 8$ for the *Rs. molischianum* LH-II ring. $E_0$ is the $Q_y$ excitation energy of the individual BChls and $V_0$ the interaction energy between neighboring BChls. For $V_0 > 0$, the state of the lowest energy is

$$E_N = E_0 - 2V_0 \tag{2}$$

Only the second and third lowest electronic excitations of the ring aggregate carry oscillator strength with respective energies given by

$$E_-^+ = E_0 - 2V_0 \cos\frac{\pi}{N} \tag{3}$$

corresponding to $n = N - 1, -N + 1$ respectively. Therefore we will focus only on (2) and (3) energies to present a four qubits Hamiltonian of the *Rs. molischianum* LH-II ring that yields these energies.

**(2) Methods**

The Hamiltonian is inspired by Feynman's Hamiltonian for a spin wave in a lattice [3]. Consider a closed loop lattice, the i-th member of which is a qubit $q_i$. The interaction between qubits is due to the exchange effect, and the Hamiltonian is given by

$$H = e + v\sum_{k=0}^{2n-1} p^{k,k+1} \tag{4}$$

where $2n$ is the total number of qubits in the lattice and $p^{k,k+1}$ are the qubits exchange operators. $e$ and $v$ are constants to be determined later. Since the lattice is a closed loop $p^{2n-1,2n} = p^{2n-1,0}$. For the *Rs. molischianum* LH-II ring the Hamiltonian (4) is explicitly

$$H = e + v\sum_{k=0}^{3} p^{k,k+1} = a + v(p^{0,1} + p^{1,2} + p^{2,3} + p^{3,0}) \tag{5}$$

Let $|\psi>$ denote a $2n$ qubits state

$$|\psi> = |q_0 q_1 \cdots q_{2n-1}> \tag{6}$$

where $q_i$ is 1 or 0 and $i = 0, 1, \ldots, 2n-1$. In the *Rs. molischianum* LH-II ring, for example, $|\psi> = |1000>$ is the state where the first qubit is 1 and all other qubits are 0. Qubits exchange operators $p^{k,k+1}$ are defined as follows

$$p^{k,k+1}|q_0 q_1 \cdots q_k q_{k+1} \cdots q_{2n-1}> = |q_0 q_1 \cdots q_{k+1} q_k \cdots q_{2n-1}> \tag{7}$$

To illustrate a point, $p^{0,1}|q_0 q_1 q_2 q_3> = |q_1 q_0 q_2 q_3>$.
To solve Schrodinger's equation with the above Hamiltonian (4) we must find eigenstates and eigenvalues of $H$. To do so, notice first that Eq. (7) implies, that $H$ operating on a state with a fixed number of 1's qubits gives a state with the same number of 1's qubits. Since all states can be represented as linear combinations of those with a fixed number of 1's qubits, we have a method of partially diagonalizing $H$. Let

$$|\psi_l> = \sum_{m=0}^{2n-1} e^{im\delta_l}|Bin(2^{2n-1-m})> \tag{8}$$

where $|Bin(2^{2n-1-m})>$ state is in the binary representation $Bin(2^{2n-1-m}) = k_1k_2\dots k_{2n}$.

More explicitly, $2^{2n-1-m} = k_1 2^{2n-1} + k_2 2^{2n-2} + \dots + k_{2n} 2^0$.Obviously, only $k_{m+1}$ is equal to 1 while all other k's are zero. $e^{i\delta_l}$ is a phase factor to be determined below. For the *Rs. molischianum* LH-II ring, $|\psi_l>$ is given as

$$|\psi_l>= \sum_{m=0}^{3} e^{im\delta_l}|Bin(2^{3-m})>= \\ =|1000>+e^{i\delta_l}|0100>+e^{2i\delta_l}|0010>+e^{3i\delta_l}|0001> \tag{9}$$

To solve the equation $H|\psi_l>= E_l|\psi_l>$,

$$e|\psi_l>+v\sum_{k=0}^{2n-1} p^{k,k+1}|\psi_l>= \\ e\sum_{m=0}^{2n-1} e^{im\delta_l}|Bin(2^{2n-1-m})>+v\sum_{m=0}^{2n-1} e^{im\delta_l}(\sum_{k=0}^{2n-1} p^{k,k+1}|Bin(2^{2n-1-m})> \tag{10}$$

We are looking for such a $H$ which yields energies (2) and (3). This requirement completely determines $e$ and $v$. All the solution details are in the Appendix A.

**(3) Results**

In the Appendix A we have obtained the four qubits Hamiltonian of the *Rs. molischianum* LH-II ring that yields energies (2) and (3)

$$H = (E_0 - 2V_0) + V_0(1-\cos\frac{\pi}{8})\sum_{k=0}^{3} p^{k,k+1}.$$

The Hamiltonian generates only 3 equidistant energy levels given by

$$E_2^H = E_0 - 2V_0$$

$$E_{1,-1}^H = E_0 - 2V_0 + 2V_0(1-\cos\frac{\pi}{8})$$

$$E_0^H = E_0 - 2V_0 + 4V_0(1-\cos\frac{\pi}{8})$$

with normalized wave functions

$$|\psi_l> = \frac{1}{2}(|1000> + e^{i\delta_l}|0100> + e^{2i\delta_l}|0010> + e^{3i\delta_l}|0001>) \quad (11)$$

where $l = -1, 0, 1, 2$ and $\delta_l = \frac{l\pi}{2}$. The energy difference between subsequent levels is given by $2V_0(1 - \cos\frac{\pi}{8})$ which is equal to 422 cm$^{-1}$ [4]. It is instructive to express the state $|\psi_l>$ in a representation which displays explicitly the entanglement (Appendix B). Hence, as we showed in the Appendix B if we subsequently divide qubits of *Rs. molischianum* LH-II ring into the subsystem A consisting of only one qubit and the subsystem B consisting of the remaining qubits, respective entropies of entanglement increase until the value of 1 for the maximally entangled state is reached.

Since the above Hamiltonian in essence introduces a two-level approximation for the LH-II ring, we will go one step further and assume that interactions between qubits closed loop lattices and an electromagnetic field are described by the Jaynes-Cummings Hamiltonian [5]

$$H_{JC} = h\nu N + \delta Z + g(a^+\sigma_- + a\sigma_+) \quad (12)$$

where $N = a^+a + \frac{Z}{2}$, $Z$ is a Pauli operator and $[H_{JC}, N] = 0$. $a^+$, $a$ are raising and lowering operators on the single mode field, $\nu$ is the frequency of the field, $\delta = \pi(\nu_0 - \nu)$ is known as the detuning where $\nu_0$ is the frequency of the qubits closed loop lattice, and $g$ is the coupling constant for the interaction between the qubits closed loop lattice and field. $\delta Z$ term originates from the two-level approximation for the LH-II ring. The Hamiltonian of the qubits closed loop lattice itself, in this two-level subspace, is $\frac{h\nu_0}{2}Z$ where $h\nu_0$ is the difference of the energies of the two levels $E_{1,-1}$ and $E_2$ since the states $|\psi_{1,-1}>$, $|\psi_2>$ are energy eigenstates. It is easy to verify that eigenstates and eigenvalues of the above Hamiltonian (12) are

$$|\psi_n> = \frac{1}{\sqrt{1+\beta_n^2}}(|n,1> + \beta_n|n+1,0>)$$
$$E_n = h\nu(n + \frac{1}{2}) + \sqrt{\delta^2 + g^2(n+1)} \quad (13)$$
$$\beta_n = \frac{\sqrt{\delta^2 + g^2(n+1)} - \delta}{g\sqrt{n+1}}$$

and

$$|\overline{\psi}_n> = \frac{1}{\sqrt{1+\overline{\beta}_n^2}}(|n,1> - \overline{\beta}_n|n+1,0>)$$

$$\overline{E}_n = h\nu(n+\frac{1}{2}) - \sqrt{\delta^2 + g^2(n+1)} \tag{14}$$

$$\overline{\beta}_n = \frac{\sqrt{\delta^2 + g^2(n+1)} + \delta}{g\sqrt{n+1}}$$

where $|0>=|\psi_2>$ and $|1>$ is, in general, a linear superposition of $|\psi_1>$ and $|\psi_{-1}>$ and $n = 0, 1, 2, \ldots$. Hence using the Schmidt's expansion the von Neuman entropies of the field $(S_F)$ or qubits closed loop lattice $(S_q)$ are given by

$$S_F = S_q = -p\log_2 p - (1-p)\log_2(1-p) \tag{15}$$

where $p = \frac{1}{1+\beta_n^2}$ or $p = \frac{1}{1+\overline{\beta}_n^2}$ depending on the eigenstate of the field. The entropies (15) reach the maximum value of 1 when $\beta_n = \overline{\beta}_n = 1$ which in essence means that the frequencies of the field and the qubits closed loop lattice are the same ( $\delta = 0$ ). It is interesting to notice that both $\beta_n$ and $\overline{\beta}_n$ approach 1 when $n \to \infty$ . From eigenvalues expressions $E_n$ and $\overline{E}_n$ we see that the qubits lattice and field oscillate back and fourth exchanging a quantum of energy, at the Rabbi frequency $\sqrt{\delta^2 + g^2(n+1)}$ which is simply $g\sqrt{n+1}$ for the case when $\delta = 0$ . Since in this case $S_F = S_q = 1$ , an equivalent description is that the qubits lattice and field oscillate back and fourth exchanging one qubit at the frequency $g\sqrt{n+1}$ .

**(4) Discussion**

Purple bacteria developed their harvesting light virtuosity in a harsh habitat - at the bottom of ponds or in topsoil, depending on the species. Only light left unharvested by plants penetrates to these depths. In essence purple bacteria must capture every single photon that comes to their habitat. Therefore, the most important question which the bacteria has to answer is how to capture a photon. Because the rules that govern the behavior of atoms and electrons and photons are very different from the standard Newtonian, classical, rules, the information that is carried by a quantum object like a photon is different from simple bits that can be inscribed on a classical object. In classical information theory, the answer to any yes/no question can always be answered as yes or no - a 1 or a 0. But with quantum theory, this nice, easy distinction between yes and no breaks down. While a classical object can never be in an ambiguous superposition of two states - it must always be in one state or another, a 1 or a 0, but not both at the same time - a quantum object can be. Mere 1s an 0s cannot capture superposition. The realm of the photons does not have the neat dichotomies of the classical world. Hence, purple bacteria utilizes quantum information.

Photosynthesis is initiated by electronic excitation of an aggregate of light-harvesting antenna complexes and by transfer of the excitation to the reaction center (RC). This organization in which multiple light-harvesting antennas (LH-II and LH-I) serve the RC is the purple bacteria's answer to the question how to capture a photon. With it, photons are captured from a broader spectral range and energy is used much more efficiently. Quantum theory deals with the transfer of quantum information. Hence, when a photon is absorbed by a LH-II complex, the quantum information from the environment (electromagnetic field) is transferred to the complex. According to our simple Hamiltonian (A13), this information transfer results in a simple change of the phase factor $\delta_l$. In fact, for the lowest *Rs. molischianum* LH-II ring state the phase factor is 0, while for the second and third lowest electronic excitations of the ring aggregate the phase factor is $\pm \frac{\pi}{2}$.

Let's start the *Rs. molischianum* LH-II ring in the lowest (ground) state and bathe it in light made of photons whose energy is equal to the difference in energy between the ring's ground $|0\rangle = |\psi_2\rangle$ state and first excited state $|1\rangle$ (in general $|1\rangle$ is a linear superposition of $|\psi_1\rangle$ and $|\psi_{-1}\rangle$ states). This corresponds to $\delta = 0$ for the Hamiltonian (12). During the jump, the ring and the light are in a state of superposition of the ground state of the ring, with no photon absorbed from the bath, and the first excited state of the ring, with one photon absorbed. Just after the ring starts its jump, the superposition is made up mostly of the ground state, with only a little bit of the excited state mixed in. Halfway through the jump, the ring and bath are in the equal superposition state |ground state, no photon absorbed> + |first excited state, photon absorbed>. Near the end of the jump, the superposition is mostly the first excited state, with only a little bit of the ring ground state remaining. The ring does not instantaneously jump from its ground state to its excited state. Rather, it glides through a continuous intermediate sequence of superpositions. At the end of the jump, the superposition consists only of the first excited state and none of the ring ground state remaining. According to (15) the ring's entropy $S_q = 1$, corresponding to the quantum information transfer of one qubit from the field to the ring.

A novel feature of the Hamiltonian (15) is an additional energy level ($E_0^H$) not found in the exciton state energies (1) since the following equation does not have an integer solution

$$E_0^H = E_0 + 2V_0 \cos\frac{n\pi}{8} .$$

Approximate solution $n \approx 6.5$ implies that $E_0^H$ is located between $E_6$ and $E_7$ energy levels of the set (1). It is 422 cm$^{-1}$ higher than $E_{1,-1}^H$.

**(5) Conclusions**

We have constructed a simple four qubits Hamiltonian of the *Rs. molischianum* purple bacteria light harvesting complex II (LH-II) ring which yields energy levels that carry the ring's oscillator strength. We have assumed that qubits form a closed loop lattice and the interaction between qubits is only due to the exchange effect. As we showed, eigenstates are constructed in a such way that as we subsequently divide qubits of the *Rs. molischianum* LH-II ring into the subsystem A consisting of only one qubit and the subsystem B consisting of the remaining qubits, respective entropies of entanglement increase until the value of 1 for the maximally entangled state (bipartite system) is reached. Since the Hamiltonian in essence introduced a two-level approximation for the LH-II ring, we went one step further and assumed that interactions between qubits closed loop lattices and an electromagnetic field are described by the Jaynes-Cummings Hamiltonian. This assumption leads to behavior where the qubits lattice and field oscillate back and fourth exchanging a quantum of energy at the Rabbi frequency or in an equivalent description the qubits lattice and field oscillate back and fourth exchanging one qubit at that frequency. This information transfer results in a simple change of the phase factor $\delta_l$ of the qubits lattice eigenstates. Thus the above assumption opens a challenging opportunity to experimentally study the *Rs. molischianum* LH-II ring in the regime of cavity QED. In addition, the Hamiltonian has an additional energy level $E_0^H$ not found in the exciton state energies . It is to be hoped that some inquirer may soon succeed in bringing in an experimental verdict for this energy level.

**Appendix A**

Using (7) we obtain

$$\begin{aligned} &p^{k,k+1}|Bin(^{2n-1-m})>= \\ &p^{m,m+1}|Bin(2^{2n-1-m})> \ when\ k = m, \\ &p^{m-1,m}|Bin(2^{2n-1-m})> \ when\ k = m-1, \\ &|Bin(2^{2n-1-m})> \ otherwise. \end{aligned} \qquad \text{(A1)}$$

It follows from (A1) that the right side of (10) is

$$\begin{aligned} &e\sum_{m=0}^{2n-1} e^{im\delta_l}\,|Bin(2^{2n-1-m})> + \\ &v\sum_{m=0}^{2n-1} e^{im\delta_l}\,(p^{m,m+1} + p^{m-1,m})|Bin(2^{2n-1-m})> + \\ &v(2n-2)\sum_{m=0}^{2n-1} e^{im\delta_l}\,|Bin(2^{2n-1-m})> \end{aligned} \qquad \text{(A2)}$$

Now

$$\begin{aligned} &p^{m,m+1}|Bin(2^{2n-1-m})>= p^{m,m+1}|00\ldots010\ldots0>= \\ &|00\ldots0010\ldots0>=|Bin(2^{2n-1-m-1})> \end{aligned}$$

where $|Bin(2^{2n-1-m})>$ and $|Bin(2^{2n-1-m-1})>$ have 1 at m-th and (m+1)-th positions respectively. Similarly,

$$\begin{aligned} &p^{m-1,m}|Bin(2^{2n-1-m})>= p^{m-1,m}|00\ldots0100>= \\ &|00\ldots100\ldots0>=|Bin(2^{2n-1-m+1})> \end{aligned}$$

Hence (A2) becomes

$$e\sum_{m=0}^{2n-1} e^{im\delta_l} \,|\, Bin(2^{2n-1-m}) > +$$
$$v\sum_{m=0}^{2n-1} e^{im\delta_l} (|\, Bin(2^{2n-1-m-1}) > + |\, Bin(2^{2n-1-m+1}) >) + \qquad (A3)$$
$$v(2n-2)\sum_{m=0}^{2n-1} e^{im\delta_l} \,|\, Bin(2^{2n-1-m}) >$$

Expanding the second sum of (A3) we obtain

$$v(|\, Bin(2^{2n-2}) > + |\, Bin(2^{2n}) > +$$
$$v\sum_{m=1}^{2n-2} e^{im\delta_l} (|\, Bin(2^{2n-1-m-1}) > + |\, Bin(2^{2n-1-m+1}) > + \qquad (A4)$$
$$ve^{i(2n-1)\delta_l} (|\, Bin(2^{-1}) > + |\, Bin(2^{1}) >)$$

A simple change of indices in (A4) results in

$$v\sum_{m=2}^{2n-1} e^{i(m-1)\delta_l} \,|\, Bin(2^{2n-1-m}) > +$$
$$\qquad (A5)$$
$$v\sum_{m=0}^{2n-3} e^{i(m+1)\delta_l} \,|\, Bin(2^{2n-1-m}) >$$

Now the first sum in (A5) for $m = 1$ gives $v|\, Bin(2^{2n-2}) >$.Similarly, the second sum in (A5) for $m = 2n-2$ gives $ve^{i(2n-1)\delta_l} \,|\, Bin(2^{1}) >$ . Thus for (A4) we obtain

$$v|\, Bin(2^{2n}) > + ve^{i(2n-1)\delta_l} \,|\, Bin(2^{-1}) > +$$
$$v\sum_{m=1}^{2n-1} e^{i(m-1)\delta_l} \,|\, Bin(2^{2n-1-m}) > + \qquad (A6)$$
$$v\sum_{m=0}^{2n-2} e^{i(m+1)\delta_l} \,|\, Bin(2^{2n-1-m}) >$$

Since the lattice is a closed loop $e^{i(-1)\delta_l} = e^{i(2n-1)\delta_l}, |\, Bin(2^{2n-1}) >= |\, Bin(2^{-1}) >$ . By the same token, $e^{i(2n)\delta_l} = 1, |\, Bin(2^{0}) >= |\, Bin(2^{2n}) >$. Hence (A6) becomes

$$v\sum_{m=0}^{2n-1} e^{i(m-1)\delta_l}|Bin(2^{2n-1-m})> + v\sum_{m=0}^{2n-1} e^{i(m+1)\delta_l}|Bin(2^{2n-1-m})> =$$
$$v\sum_{m=0}^{2n-1} e^{im\delta_l}\, 2\cos(\delta_l)|Bin(2^{2n-1-m})> \quad \text{(A7)}$$

Therefore $H|\psi_l>= \sum_{m=0}^{2n-1}[e + 2v\cos(\delta_l) + v(2n-2)]e^{im\delta_l}|Bin(2^{2n-1-m})>$. Since $E_l|\psi_l>= E_l\sum_{m=0}^{2n-1} e^{im\delta_l}|Bin(2^{2n-1-m})>$ we obtain for eigenvalues

$$E_l = e + 2v\cos(\delta_l) + v(2n-2) \quad \text{(A8)}$$

The lattice is a closed loop, therefore
$e^{2in\delta_l} = 1$ implies $\delta_l = \frac{l\pi}{n}$ where $l = -n+1, -n+2, \ldots, n-1, n$ .Hence we obtain for (A8)

$$E_l = e + 2v\cos(\frac{l\pi}{n}) + v(2n-2) \quad \text{(A9)}$$

Finally, for the *Rs. molischianum* LH-II ring,

$$E_l = e + 2v\cos(\frac{l\pi}{2}) + 2v \quad \text{(A10)}$$

and $l = -1, 0, 1, 2$ . Explicitly,

$$E_2 = e$$
$$E_{1,-1} = e + 2v \quad \text{(A11)}$$
$$E_0 = e + 4v$$

We demand $E_2 = E_N$ and $E_{1,-1} = E_-^+$ where $N = 8$ for the *Rs. molischianum* LH-II ring. Using (2) and (3) we arrive at

$$e = E_0 - 2V_0$$
$$v = V_0(1 - \cos\frac{\pi}{8}) \quad \text{(A12)}$$

In order to avoid confusion with $E_0$ which represents the $Q_y$ excitation energy of the individual BChls, we will label energy levels (A11) as $E_l^H$. Finally, if we substitute the above values for $\varepsilon$ and $V$ into (5) we obtain the four qubits Hamiltonian of the *Rs. molischianum* LH-II ring that yields energies (2) and (3)

$$H = (E_0 - 2V_0) + V_0(1 - \cos\frac{\pi}{8})\sum_{k=0}^{3} p^{k,k+1} \qquad \text{(A13)}$$

**Appendix B**

In general, the superposition in (11) does not tell us at first sight whether the state can be factored. The quantum content of the global state ($|\psi_l>$) is intricately interwoven between the parts (qubits). In many situations we are interested in performing manipulations or measurements on one part without looking at the other. For example, we might wish to know the probability of finding a given result when measuring an observable $O_1$ attached to the qubit 1 (the subsystem A), without worrying about what happens to the rest of the qubits (the qubits 2, 3, and 4 are the subsystem B). Predictions about the outcome of such an experiment can be made by using the density operator formalism.
The density operator $\rho_l$ of the *Rs. molischianum* LH-II ring described by the quantum state $|\psi_l>$ is the projector onto this state

$$\rho_l = |\psi_l >< \psi_l| . \qquad \text{(B1)}$$

This formalism becomes advantageous for describing the subsystem A without looking at the subsystem B. We can build a partial density operator $\rho_A$ containing all predictive information about A alone, by tracing $\rho_l$ over the subspace B

$$\rho_A = Tr_B(\rho_l) . \qquad \text{(B2)}$$

In fact, we can make predictions on A without having to consider B. It is important to recognize however that the information content in $\rho_A$ and in the equivalent partial density operator $\rho_B = Tr_A(\rho_l)$ is smaller than in $\rho_l$ since we renounce, by considering A alone, accounting for the correlations between A and B. Stating that A and B are entangled is equivalent to saying that the partial density operators $\rho_A$ and $\rho_B$ are not projectors on a quantum state [6]. They are, however, hermitian operators which can be expanded as a sum of projectors. There is at least a basis $B_A$ in which $\rho_A$ is diagonal (an infinity of bases if one eigenvalue is degenerate). In one of this bases, $\{|j_A>\}$, $\rho_A$ is

$$\rho_A = \sum_j \lambda_j |j_A >< j_A| , \qquad \text{(B3)}$$

where the $\lambda_j$ are positive or zero eigenvalues whose sum over $j$ is equal to 1. Equation (B3) appears as a probabilistic expansion. By overlooking what happens to B, we have only a statistical knowledge of the state of A, with a probability $\lambda_j$ of finding it in $|j_A>$. The expansion (B3) is quite different from a quantum mechanical superposition of states. Here the coefficients are positive probabilities and not complex amplitudes. If A and B are not entangled, one $\lambda_j$ is equal to 1 and all others are zero. As soon as at least two $\lambda_j$'s are non-zero, A and B are entangled.
It is instructive to write equation (11) as

$$\psi_l = \sum_{i,k} \alpha_{ik} |i_A>|k_B> . \qquad \text{(B4)}$$

Then choosing a basis $B_A$ in which $\rho_A$ is diagonal, (B4) can be expressed as

$$|\psi_l>= \sum_j |j_A>|\tilde{j}_B> \qquad \text{(B5)}$$

where $|\tilde{j}_B>= \sum_k \alpha_{jk}|k_B>$ . $|\tilde{j}_B>$ states are orthogonal to each other, since

$$<j_A|\rho_A|j'_A>= \lambda_j \delta_{jj'} =<\tilde{j}_B|\tilde{j}'_B> .$$

Finally, normalizing states of the subsystem B by the transformation $|\tilde{j}_B>=|\tilde{j}_B>/\sqrt{\lambda_j}$ , we obtain the Schmidt expansion

$$|\psi_l>= \sum_j \sqrt{\lambda_j}|j_A>|\tilde{j}_B> \qquad \text{(B6)}$$

Equation (B6) exhibits in a transparent way the entanglement between A and B.
Let us now derive (B6) explicitly for the *Rs. molischianum* LH-II ring. Substituting (B1) into (B2) and using equation (11) we obtain the following expression for $\rho_A$:

$$\rho_A = \frac{1}{4}|1_A><1_A|+\frac{3}{4}|0_A><0_A| . \qquad \text{(B7)}$$

Hence

$$|\tilde{1}_B>= \frac{1}{2}|000>$$

$$|\tilde{0}_B>= \frac{1}{2}(e^{i\delta_l}|100>+e^{2i\delta_l}|010>+e^{33i\delta_l}|001>)$$

It is easy to verify

$$< 1_A|\rho_A|1_A> = < 1_{\tilde{B}}|1_{\tilde{B}}> = \frac{1}{4}$$

$$< 0_A|\rho_A|0_A> = < 0_{\tilde{B}}|0_{\tilde{B}}> = \frac{3}{4}.$$

Thus we obtain for equation (B6)

$$|\psi_l> = \sqrt{\frac{1}{4}}|1>|000> + \sqrt{\frac{3}{4}}|0> \frac{1}{\sqrt{3}}(e^{i\delta_l}|100> + e^{2i\delta_l}|010> + e^{3i\delta_l}|001>) \quad \text{(B8)}$$

with the normalized states of the subsystem B given by

$$|1_{\tilde{B}}> = |000>$$

$$|0_{\tilde{B}}> = \frac{1}{\sqrt{3}}(e^{i\delta_l}|000> + e^{2i\delta_l}|010> + e^{3i\delta_l}|001>)$$

We will now focus on a measure of the degree of entanglement between the subsystems A (the qubit 1) and B (the qubits 2, 3, and 4). It is intuitive that the more the $\lambda_j$ probabilities are spread out over many non-zero values, the more information we lose by focusing separately on one subsystem, disregarding the correlations between A and B. It is reasonable to link this loss of mutual information to the degree of entanglement. A natural way to measure this loss of information is to compute the von Neuman entropy of A or B, defined as

$$S_A = S_B = -\sum_j \lambda_j \log_2 \lambda_j = = -Tr(\rho_A \log_2 \rho_A) = -Tr(\rho_B \log_2 \rho_B) \quad \text{(B9)}$$

The entropy of entanglement $S_e = S_A = S_B$ express quantitatively the degree of disorder in our knowledge of the partial density matrices of the two parts of the entangled system. For the above subsystems A and B of the *Rs. molischianum* LH-II ring

$$S_e = -(\frac{1}{4}\log_2\frac{1}{4} + \frac{3}{4}\log_2\frac{3}{4}) = = -(-\frac{1}{2} + \frac{3}{4}\log_2 3 - \frac{3}{2}) = 2 - \frac{3}{4}\log_2 3 \approx 0.415 \quad \text{(B10)}$$

In a maximally entangled state for a bipartite system, local measurements performed on one part of the system are not predictable at all. For a such state $S_e = 1$ and $\lambda_j = \frac{1}{2}$. It is obvious from equation (B10) that the above subsystems A and B of the *Rs. molischianum* LH-II ring are not maximally entangled, but they are fairly close to such a state.
Let

$$|\psi_l^{'}>= \frac{e^{i\delta_l}}{\sqrt{3}}(|100> +e^{i\delta_l}|010> +e^{2i\delta_l}|001>) \quad . \qquad \text{(B11)}$$

Hence

$$|\psi_l>= \sqrt{\frac{1}{4}}|1>|000> + \sqrt{\frac{3}{4}}|0>|\psi_l^{'}> .$$

It is interesting to obtain the Schmidt expansion for $|\psi_l^{'}>$. Repeating relevant steps above we arrive at

$$|\psi_l^{'}>= \sqrt{\frac{1}{3}}|1> e^{i\delta_l}|00> + \sqrt{\frac{2}{3}}|0> \frac{e^{2i\delta_l}}{\sqrt{2}}(|10> +e^{i\delta_l}|01>) \qquad \text{(B12)}$$

Then the entropy of entanglement between the subsystem A (the second qubit) and subsystem B ( the third and fourth qubits) is given by

$$\begin{aligned} S_e &= -(\frac{1}{3}\log_2\frac{1}{3} + \frac{2}{3}\log_2\frac{2}{3}) = \\ &= -\frac{2}{3} + \log_2 3 = 0.918 \end{aligned} \qquad \text{(B13)}$$

Let now

$$|\psi_l^{''}>= \frac{e^{2i\delta_l}}{\sqrt{2}}(|10> +e^{i\delta_l}|01>) \quad . \qquad \text{(B14)}$$

Hence

$$|\psi_l^{'}>= \sqrt{\frac{1}{3}}|1> e^{i\delta_l}|00> + \sqrt{\frac{2}{3}}|0>|\psi_l^{''}>$$

and the Schmidt expansion for the third (the subsystem A) and fourth (the subsystem B) qubits is

$$|\psi_i^{''}> = \sqrt{\frac{1}{2}}|1> e^{2i\delta_i}|0> + \sqrt{\frac{1}{2}}|0> e^{3i\delta_i}|1>.$$

The entropy of entanglement in this case is

$$S_e = -(\frac{1}{2}\log_2\frac{1}{2} + \frac{1}{2}\log_2\frac{1}{2}) = 1.$$

The third and fourth qubits are maximally entangled. Therefore, as we subsequently divide qubits of the *Rs. molischianum* LH-II ring into the subsystem A consisting of only one qubit and the subsystem B consisting of the remaining qubits, respective entropies of entanglement increase until the value of 1 for the maximally entangled state (bipartite system) is reached.

**Acknowledgments**
The author would like to thank to Professor Marilyn Gunner, City College of New York for asking inspiring questions which initiated this paper.